\documentclass[aps,prd,showkeys,nofootinbib,twocolumn,superscriptaddress]{revtex4-2}
\usepackage{enumerate}
\usepackage{amsmath,slashed,tensor,amssymb,epsfig,amsthm,bm,graphicx,graphics,slashed}
\usepackage[caption=false]{subfig}
\usepackage{hyperref}
\usepackage[utf8]{inputenc}
\usepackage[top=1.5cm, bottom=1.5cm, left=2cm, right=2cm]{geometry}

\usepackage{makecell}
\usepackage{microtype}
\usepackage[font=footnotesize,labelfont=bf]{caption}

\newcommand{\be}{\begin{equation}}\newcommand{\ee}{\end{equation}}
\newcommand{\bea}{\begin{eqnarray}}\newcommand{\eea}{\end{eqnarray}}
\newcommand{\brr}{\begin{array}}\newcommand{\err}{\end{array}}
\newcommand{\bit}{\begin{itemize}}\newcommand{\eit}{\end{itemize}}
\newcommand{\ben}{\begin{enumerate}}\newcommand{\een}{\end{enumerate}}

\newcommand{\bbm}{\begin{bmatrix}}\newcommand{\ebm}{\end{bmatrix}}
\newcommand{\ba}{\begin{array}}
\newcommand{\ea}{\end{array}}

\newtheorem{mydef}{Definition}
\newtheorem{Lemma}{Lemma}
\newtheorem{theorem}{Theorem}
\newcommand{\bd}{\begin{mydef}} \newcommand{\ed}{\end{mydef}}
\newcommand{\bthe}{\begin{theorem}} \newcommand{\ethe}{\end{theorem}}
\newcommand{\ble}{\begin{Lemma}} \newcommand{\ele}{\end{Lemma}}

\newcommand{\dr}{\mathrm{d}}

\def\ha{\frac{1}{2}}

\def\lan{\langle}
\def\lf{\left}

\def\non{\nonumber}\def\pa{\partial}\def\ran{\rangle}

\def\ri{\right}

\def\de{\delta}\def\De{\Delta}

\def\la{\lambda}\def\La{\Lambda}

\def\eps{\varepsilon}

\def\1{{_{1}}}\def\2{{_{2}}}

\newcommand{\ide}{1\hspace{-1mm}{\rm I}}

\def\noHe0{:\;\!\!\;\!\!:H_e(0):\;\!\!\;\!\!:}
\def\noHm0{:\;\!\!\;\!\!:H_\mu(0):\;\!\!\;\!\!:}

\def\lan{\langle}
\def\lf{\left}

\def\non{\nonumber}
\def\pa{\partial}\def\ran{\rangle}

\def\ri{\right}

\def\de{\delta}\def\De{\Delta}

\def\la{\lambda}
\def\La{\Lambda}

\def\1{{_{1}}}\def\2{{_{2}}}

\begin{document}

%%%%%%%%%%%%%%%%%%%%%%%%%%%%%%%%%%%%%%%%%%%%%%%%%%%%%%%%%%%%%%%%%%%%%%%%%%%%%%%%%%%%%%%%%%%%
\title{Quantum Groups and Polymer Quantum Mechanics}
%%%%%%%%%%%%%%%%%%%%%%%%%%%%%%%%%%%%%%%%%%%%%%%%%%%%%%%%%%%%%%%%%%%%%%%%%%%%%%%%%%%%%%%%%%%%
%%%%%%%%%%%%%%%%%%%%%%%%%%%%%%%%%%%%%%%%%%%%%%%%%%%%%%%%%%%%%%%%%%%%%%%%%%%%%%%%%%%%%%%%%%%%
\author{G.~Acquaviva}
\email{gioacqua@gmail.com}

\affiliation{Arquimea Research Center, Camino de las Mantecas, 38320, Santa Cruz de Tenerife, Spain}

\author{A.~Iorio}
\email{iorio@ipnp.troja.mff.cuni.cz}

\affiliation{Institute of Particle and Nuclear Physics, Faculty  of  Mathematics  and  Physics, Charles  University, V  Hole\v{s}ovi\v{c}k\'{a}ch  2, 18000  Praha  8,  Czech  Republic}

\author{L.~Smaldone}
\email{smaldone@ipnp.mff.cuni.cz}

\affiliation{Institute of Particle and Nuclear Physics, Faculty  of  Mathematics  and  Physics, Charles  University, V  Hole\v{s}ovi\v{c}k\'{a}ch  2, 18000  Praha  8,  Czech  Republic}

\begin{abstract}
In Polymer Quantum Mechanics, a quantization scheme that naturally emerges from Loop Quantum Gravity, position and momentum operators cannot be both well-defined on the Hilbert space ($\mathcal{H}_{Poly}$). It is henceforth deemed impossible to define standard creation and annihilation operators. In this letter we show that a $q$-oscillator structure, and hence $q$-deformed creation/annihilation operators, can be naturally defined on $\mathcal{H}_{Poly}$, which is then mapped into the sum of many copies of the $q$-oscillator Hilbert space. This shows that the $q$-calculus is a natural calculus for Polymer Quantum Mechanics. Moreover, we show that the inequivalence of different superselected sectors of $\mathcal{H}_{Poly}$ is of topological nature.
%%%%%%%%%%%%%%%%%%%%%%%%%%%%%%%%%%%%%%%%%%%%%%%%%%%%%%%%%
\end{abstract}

\vspace{-1mm}

\maketitle
%
%%%%%%%%%%%%%%%%%%%%%%%%%%%%%%%%%%%%%%%%%%%%%%%%%%%%%%%%%%%%%%%%%%%%%%%
\section{Introduction}

Quantum groups are rich mathematical structures, born in the 1980s through the work of Faddeev and the Leningrad school, originally in the context of integrable systems \cite{Faddeev:1982rn}. Nowadays, their applications in physics are as widespread as those of Lie algebras, see, e.g., \cite{castellani1996quantum}, \cite{majid2000foundations}, \cite{manin2018quantum}. Indeed, they can be referred to as Hopf algebras that, in many cases, are one parameter ($q$) deformations of the universal envelope of Lie algebras. This mathematical terminology, although precise, does not do justice to neither words of the expression ``quantum group'', that is customarily used, and that points to the most ambitious part of that research programme: to provide the mathematical structures to handle the full, hence \emph{exact}, quantization of systems, as opposed to the infinitesimal, hence approximated, standard quantization.

In Manin's original language \cite{manin1988Montreal}, we have the quantum group $GL(2)_q$ as symmetry of the full quantum phase-space, whose variables satisfy
\be \label{gl2q}
x p = e^\hbar p x \,,
\ee
with the deformation parameter $q \equiv e^\hbar$, whereas the classical phase-space has $GL(2)$ as symmetry group. Hence, in this programme, $q = 1 + \hbar + \cdots$ should correspond to $q = $``classical'' + ``standard (approximated) quantum'' + ``full (exact) quantum''. As such, this did not find a satisfactory and universal implementation as a fundamental physics principle of nature.

In this letter we consider another road to give quantum groups a fundamental meaning: that is, the road of quantum gravity, where a fundamental scale, related to the parameter $q$, emerges.  Hence quantum groups might reveal to be the most fundamental algebraic quantum structures of nature, see, e.g., \cite{Majid_1988} and \cite{majid2000foundations}.
This is the point of view of Ref. \cite{Bianchi:2011uq}, where it is argued that, in the presence of a minimal length, such as the Planck length $\ell_{_{P}}$, and of a cosmological constant $\La$, the appropriate ``rotation group'' is $SU(2)_q$, with  $q = \exp\{ i \ell_{_{P}}^2 \Lambda \} $. On a similar line of reasoning, $q$-deformed formulations of Loop Quantum Gravity (LQG) were proposed along the years \cite{Major_1996, Noui:2002ag, Fairbairn:2010cp, Dittrich:2013voa}, which naturally incorporate the cosmological constant. This relationship was even used in different models \cite{Jalalzadeh:2017jdo}. It has been also shown that the quantum group structure of the $\kappa$-Poincar\'e is the natural symmetry for effective field theories, emerging from discrete gravity theories coupled to matter \cite{Freidel:2005me}, and for noncommutative description of the spacetime
\cite{doi:10.1063/1.533331, PhysRevD.99.085003, lizzi2021kappapoincarecomodules}. For other approaches see also \cite{IORIO2001234}.

Despite the rich literature, the general issue remains still completely open, and part of the focus is shifting on the so-called \emph{polymer quantization} (PQ) \cite{Ashtekar:2001xp,Ashtekar:2002vh} (that is a slightly simplified quantization inspired by LQG, that is reliable and heavily used \cite{Ashtekar:2011ni,Hassan:2014sja,Bonder:2017ckx,Stargen:2019pme,Escobar:2020zat,Blanchette:2020kkk,Garcia-Chung:2020zyq}) and to its application to finite degrees of freedom systems, \emph{polymer quantum mechanics} (PQM) \cite{Ashtekar:2002sn,PhysRevD.76.044016,Morales_T_cotl_2017}. PQ is based on the \emph{polymer representation} of the Weyl--Heisenberg (WH) algebra, which is a non-regular representation, inequivalent to the standard Schr\"odinger or Fock--Bargmann representations \cite{woit2017quantum}. In PQM
%(in the so-called $q$-polarization)
position operator has discrete eigenvalues, while momentum operator is not well defined. As a consequence, only finite translations can be considered, and a lattice structure naturally emerges, while usual symmetry groups are deformed \cite{Chiou_2007, Date:2012gf, Amelino_Camelia_2017}. In Ref. \cite{Amelino_Camelia_2017} it was pointed out how discrete structure of PQM could be related to the emergence of a $\kappa$-Poincar\'e structure at the Planck-scale. Moreover, it has been shown that even standard creation and annihilation operators in PQ cannot be defined, and should be replaced by some deformed objects \cite{Husain:2010gb}.

The role of the $q$-deformed WH algebra as an essential tool in the physics of discrete quantum systems, has been extensively discussed \cite{Celeghini_1995,qSvN,Fichtmuller_1996}, and is by now well known. The emphasis there is on condensed matter systems, rather than fundamental ones. A related application was also the quantum particle moving on a circle \cite{Kowalski:1998hx}.

Here we shall build on that, and shall bring together those ideological strands, focusing on PQM, as a prototypical discrete fundamental quantum system. We show that the probably most primordial quantum group, that is the $q$-oscillator algebra \cite{Biedenharn:1989jw, Macfarlane:1989dt}, reveals to be a natural mathematical set-up for PQM. We do so by first constructing an explicit representation of the $q$-oscillator algebra on the superselected sectors of the Hilbert space of PQM. Then, we show that \emph{each such sector} can be mapped into the sum of \emph{two copies} of $q$-oscillator Hilbert space. Moreover, we show that the inequivalence of different representations of $q$-WH, corresponding to different superselected Hilbert spaces, is related to the topological inequivalent quantizations of a quantum particle moving on a circle. Finally, the role of $q$-calculus as the natural calculus for PQM is emphasized.

%In particular, we shall be able to naturally define creation and annihilation operators, that is the most importat first step to quantize a physical system.

%***CONSIDER using 5 pages***

%***CONSIDER removing the sections (one can use, e.g,, ``{\it Introducion.} Quantum groups are rich mathematical structures [...]''***

%%%%%%%%%%%%%%%%%%%%%%%%%%%%%%%%%%%%%%%%%%%%%%%%%%%%%%%%%%%%%%%%%%%%%%%

\section{Basics on PQM and on $q$-WH}
%\subsection{Polymer quantum mechanics}
The basic algebra of quantization, that is the WH algebra,
%written in the $C^*$ algebra language,
is defined by
\bea
U(\la) \, U(\nu) & = & U(\la +\nu) \, , \quad V(\mu) \, V(\rho) \ = \ V(\mu +\rho) \, , \\[2mm] \label{cr}
U(\la) \, V(\mu) & = & e^{-i \, \la \, \mu} \,V(\mu) \, U(\la) \, ,
\eea
where $U$, $V$ are bounded operators on some Hilbert space $\mathcal{H}$, $\la,\mu, \nu, \rho \in \mathbb{R}$, $U(0)=V(0)=\ide$ and $U(-\la)=U^*(\la)$, $V(-\mu)=V^*(\mu)$.

\emph{The polymer representation} of the WH algebra is a non-regular representation on the space of \emph{cylindrical functions}
\be
f(k)=\sum_n f_n e^{-i k x_n}
\ee
for all possible discrete sets $\lf\{x_n \in \mathbb{R}\ri\}$. Such Hilbert space is usually indicated as $\mathcal{H}_{Poly}=L^2(\mathbb{R}_b,\dr \mu_H)$, and it is the \emph{Bohr compactification} of the real line \cite{thiemann_2007}, equipped with a translation invariant Haar measure \cite{Ashtekar:2002vh,PhysRevD.76.044016, Berra_Montiel_2021}.

An uncountable basis is thus given by the functions $(k|x\ran \equiv e^{-i k x}$, such that $\lan x|y\ran=\de_{x \, y}$, and $\de_{x \, y}$ is to be seen as a \emph{Kronecker delta} rather than a Dirac delta.

Explicitly $U(\la) f(k) = f(k-\la)$, $V(\mu) f(k)  =   e^{i  \mu  k} f(k)$ or
$U(\la) |x_n\ran  =  e^{i  \la  x_n} \, |x_n\ran$, $V(\mu) |x_n\ran  =  |x_n-\mu\ran$.

It is also possible to define a position operator $Q \equiv \lf. -i\frac{\dr U(\la)}{\dr \la} \ri|_{\la=0}$, such that
\be
Q \, |x_n\ran \ = \ x_n \, |x_n\ran \, ,
\ee
but it is \emph{not possible} to define the corresponding momentum operator, because $V(\mu)$ is discontinuous in $\mu$. In fact $\lim_{\mu \to 0}\lan x|V(\mu)|x\ran \ = \  0$, while $\lan x|V(0)|x\ran=1$ \cite{Ashtekar:2002vh}. Then, the usual ladder operators descending from $Q\pm i P$ cannot be defined on $\mathcal{H}_{Poly}$ \cite{Husain:2010gb}.
 %*** making problematic standard procedures, such as, e.g., the second quantization of a field.***

Such discontinuity in the generator of space-translation naturally leads to the existence of a minimal length $\varepsilon$, known as \emph{polymer length}, such that $x_n=x_0+n \eps$, $x_{0} \in [0,\varepsilon)$, i.e. $\mathcal{H}_{Poly}$ decomposes into the direct sum of separable (superselected) Hilbert spaces \cite{Ashtekar:2002vh,Kunstatter_2009}
\be \label{hdec}
\mathcal{H}_{Poly} \ = \ \bigoplus_{x_0 \in [x_0,\eps)} \, \mathcal{H}^{x_0}_{Poly} \, .
\ee
and, in each $\mathcal{H}^{x_0}_{Poly}$, a state $|\psi\ran_{x_0}$ can be expanded as
\bea \label{psi2n}
|\psi\ran_{x_0} & = & \sum^{+\infty}_{n=-\infty} \, c_{n} \, |x_n \ran \ \equiv \ \sum^{+\infty}_{n=-\infty} \, c_{n} \, |x_0+n \eps\ran \, , \\[2mm]
 \label{psik0}
\psi_{x_0}(k) & = & (k|\psi\ran_{x_0} \ = \ e^{-i \, k \, x_0} \, \sum^{+\infty}_{n=-\infty} \, c_{n} \, e^{-i \, n \, k \eps} \, ,
\eea
Note that $\mathcal{H}_{Poly}$ is still non-separable.

A momentum operator on the lattice is thus defined as  \cite{Ashtekar:2002vh,Chiou_2007, Kunstatter_2009,PhysRevD.76.044016,Amelino_Camelia_2017}
\be
P_\varepsilon \ \equiv \ -i \, \frac{V(\varepsilon)-V(-\varepsilon)}{2 \, \varepsilon} \, .
\ee
By using $[Q \, , \, V(\mu)]=-\mu \, V(\mu)$, one finds \cite{Chiou_2007}:
\be \label{e21}
\lf[Q \, , \, P_\varepsilon\ri]  \ = \ i \, I_\varepsilon \, , \quad [Q, I_\eps] = - i  \eps^2 \, P_\eps\, , \quad [P_\eps, I_\eps] = 0 \, ,
\ee
where $I_\varepsilon \equiv \frac{V(\varepsilon)+V(-\varepsilon)}{2 }$. These are basic commutators of $e(2)$ Lie algebra. An explicit representation on $f(k)$ can be taken from lattice quantum mechanics (LQM) \cite{Celeghini_1995, Jizba:2009qf}:
\be \label{lqme2}
Q \ = \ i \frac{\dr}{\dr k} \, , \, P_\eps \ = \ \frac{\sin (k \varepsilon)}{\varepsilon} \, , \, I_\eps \ = \ \cos(k \eps) \, .
\ee
%
%%%%%%%%%%%%%%%%%%%%%%%%%%%%%%%%%%%%%%%%%%%%%%%%%%%%%%%%%%%%%%%%%%%%%%%%%
%\subsection{The quantum group $q$-WH}
The $q$-oscillator algebra, or $q$-WH quantum group, is defined by \cite{Macfarlane:1989dt, PhysRevLett.66.2056, klimyk2012quantum}
\bea \label{qwho}
\hspace{-0.4cm} \lf[a \, , \, a^\dag \ri]_{q}  = q^{-N} \, ,  \quad \lf[N, a^\dag\ri] = a^\dag \, , \quad \lf[N, a\ri] =  -a \, ,
\eea
where $[A,B]_q\equiv A B-q BA$, and $q \in \mathbb{C}$. By introducing the operator $H \equiv N+\ha$, the relations \eqref{qwho} lead to the graded Hopf algebra $B(0|1)$ \cite{Celeghini:1990km,Celeghini_1995}
\be  \label{hopf}
\lf\{a, a^\dag\ri\} \ = \ [2 \, H]_{\sqrt{q}} \, , \, \lf[H, a^\dag\ri] \ = \ a^\dag \, , \, \lf[H, a\ri] \ = \ -a \, ,
\ee
with the relative coproduct maps
\bea
\De(H) & = & H \otimes \ide+\ide \otimes H \,, \label{DeH} \\[2mm]
\De(a) & = & a \otimes q^{\frac{H}{2}}+ q^{-\frac{H}{2}} \otimes a \, ,\\[2mm]
 \De(a^\dag) & = & a^\dag \otimes q^{\frac{H}{2}}+ q^{-\frac{H}{2}} \otimes a^\dag \, ,
\eea
where (\ref{DeH}) implies $\De(N) \ = \ N \otimes \ide+\ide \otimes N+\ha \ide \otimes \ide$,
and where \emph{symmetric $q$-numbers} are defined as \cite{klimyk2012quantum, kac2001quantum}
\be
\lf[x\ri]_q \ \equiv  \ \frac{q^x -q^{-x}}{q-q^{-1}} \, .
\ee

Defining $b \equiv q^{-N/2} a$, $b^\dag \equiv a^\dag q^{-N/2} $, and renaming $q^{-2} \to q$, the relations \eqref{qwho} could be rewritten fully in terms of standard commutators \cite{klimyk2012quantum,Celeghini_1995,qSvN}:
\be \label{nqwhb}
\lf[b, b^\dag\ri] \ = \ q^N \, , \quad \lf[N, b^\dag\ri] \ = \ b^\dag \, , \quad \lf[N, b\ri] \ = \ -b \,.
\ee
Notice that $N$ is the same in both versions, while for the $b$-modes it is useful to introduce the {\it non-symmetric} $q$-numbers \cite{klimyk2012quantum, kac2001quantum}
\be \label{nsqn}
[\![x]\!]_q \ \equiv \ \frac{q^x-1}{q-1} \,.
\ee
In what follows we shall denote $\lf[x\ri]_q  \equiv \ \lf[x\ri]$ and $[\![x]\!]_q \equiv \ [\![x]\!]$.
%%%%%%%%%%%%%%%%%%%%%%%%%%%%%%%%%%%%%%%%%%%%%%%%%%%
\section{$q$-WH structure of PQM}
%%%%%%%%%%%%%%%%%%%%%%%%%%%%%%%%%%%%%%%%%%%%%%%%%
Let us now show the polymer representation of $q$-oscillator. We obtain:
\bea \label{apoly}
\hspace{-0.4cm} a \, f(k) & \equiv & V(\eps) \, \frac{\sin Q_{x_0}}{\sin \eps}  \, f(k)  = V(\eps)  \,  \lf[\frac{Q_{x_0}}{\eps}\ri] \, f(k)  \label{qnun} \, ,  \\[2mm]
\hspace{-0.4cm} a^\dag \, f(k)  & \equiv & V^*(\eps) \, f(k)  \, ,\ \ \ N \, f(k)   \equiv \frac{Q_{x_0}}{\eps} \, f(k) \, , \label{npoly}
\eea
with $f(k) \in \mathcal{H}_{Poly}$, $q \equiv e^{-i \eps}$, and
\be \label{qx0}
Q_{x_0} \ \equiv \ Q-x_0 \, .
\ee

On ket states:
\bea
a \, |x_n\ran & = & \frac{\sin(n \eps)}{\sin \eps} \, |x_n-\eps\ran  \ = \ [n] \, |x_n-\eps\ran\,  \, , \\[2mm]
 a^\dag \, |x_n\ran  & = & |x_n+\eps\ran  \, ,  \ \ \ N \, |x_n\ran  \ = \ n \, |x_n\ran \, .
\eea
It is easy to see that they satisfy the relations \eqref{qwho}. The coproducts become:
\bea
\hspace{-0.4cm}\De(a) & = & V(\eps)\lf[\frac{Q_{x_0}}{\eps}\ri] \otimes e^{-\frac{i}{2} \lf(Q_{x_0}+ \frac{\eps}{2}\ri)} \non \\[1mm]
\hspace{-0.4cm} & + &  e^{\frac{i}{2} \lf(Q_{x_0}+ \frac{\eps}{2}\ri)} \otimes V(\eps) \lf[\frac{Q_{x_0}}{\eps}\ri]  \, ,  \\[1mm]
\hspace{-0.4cm}\De(a^\dag) & = & \!\! V^*(\eps) \otimes e^{-\frac{i}{2} \lf(Q_{x_0}+ \frac{\eps}{2}\ri)}
 +  e^{\frac{i}{2} \lf(Q_{x_0}\! + \! \frac{\eps}{2}\ri)} \otimes V^*(\eps) \,,   \\[1mm]
\hspace{-0.4cm}\De(N) & = & \frac{1}{\eps} \, \lf(Q_{x_0} \otimes \ide+\ide \otimes Q_{x_0}\ri) \, + \, \ha \, \ide \otimes \ide \, .
\eea
We can also find a polymer representation of $q$-WH in the other form of the commutation relations \eqref{nqwhb}:
\bea \label{qo1b}
\hspace{-0.4cm} b \, f(k) & \equiv &    V(\eps) \,\lf[\!\lf[\frac{Q_{x_0}}{\eps}\ri]\!\ri] \, f(k)\, ,  \\[2mm] \label{qo2b}
\hspace{-0.4cm} b^\dag \, f(k)  & \equiv &  V^*(\eps) \, f(k)  \, , \ \ \ N \, f(k)  \equiv  \frac{Q_{x_0}}{\eps} \, f(k) \,.
\eea
% where the $q$-number was introduced in Eq. \eqref{nsqn}.
Comparing these with (\ref{qnun})-(\ref{npoly}), we see that the only difference resides in the definition of the $q$-operator/number: $[x]$ is associated with the symmetric $q$-derivative $\bar{D}_q$ (see Eq.\eqref{qfbr} below), while $[\![x]\!]$ is associated to the $q$-derivative $\bar{D}_q$ (see Eq.\eqref{qfbr1} below) \cite{klimyk2012quantum, kac2001quantum}.

Explicitly:
\bea
b \, |x_n\ran & = &  \frac{e^{-i \, n \, \eps}-1}{e^{-i \eps}-1} \, |x_n-\eps\ran \ = \ [\![n]\!] \, |x_n-\eps\ran  \, , \\[2mm]
b^\dag \, |x_n\ran  & = & |x_n+\eps\ran  \, , \ \  \ N \, |x_n\ran  =  n \, |x_n\ran \, .
\eea
It is trivial to verify that the relations \eqref{nqwhb} are satisfied.

%*** ACCENNO AI COPRODOTTI PER I MODI $b$ ***

For $n \geq 0$, we now define:
\be
|n\ran \ \equiv \ \frac{|x_n\ran}{([n]!)^\ha} \, ,
\ \ \
|n\ran \!\ran \ \equiv \ \frac{|x_n\ran}{([\![n]\!]!)^\ha} \, .
\ee
Then one can easily check that
\bea
a |0\ran & = & 0 \, , \qquad |n\ran \ = \ \frac{(a^\dag)^n}{([n]!)^\ha} \, |0\ran \, , \\[2mm]
a^\dag \, |n\ran & = & [n+1]^\ha \, |n+1\ran \, , \quad a \, |n\ran \ = \ [n]^\ha \, |n-1\ran \, , \\[2mm]
N \, |n\ran & = & n |n\ran \, .
\eea
and
\bea
b |0\ran \! \ran & = & 0 \, , \qquad |n\ran \!\ran \ = \ \frac{(b^\dag)^n}{([\![n]\!]!)^\ha} \, |0\ran \!\ran \, , \\[2mm]
b^\dag \, |n\ran \!\ran & = & [\![n+1]\!]^\ha \, |n+1\ran \!\ran \, ,  \,\,
  b \, |n\ran \!\ran =  [\![n]\!]^\ha \, |n-1 \ran \!\ran \, ,  \\[2mm]
N \, |n\ran \!\ran   & = & n |n \ran \!\ran \, .
\eea
%I
Then $|n\ran$ ( $|n\ran\!\ran$), with the scalar product $\lan n|m\ran = \de_{n \, m}$ ($\lan \!\lan n|m\ran\!\ran = \de_{n \, m}$), form a basis for the Hilbert space of $q$-oscillator $\mathcal{H}_q$ ($\mathcal{H}'_q$), originally introduced in Refs.\cite{Biedenharn:1989jw,Macfarlane:1989dt}.

For $n \leq 0$, we define $m \equiv-n \geq 0 $ and $\eta=-\eps$. Then $|x_n\ran$ $=$ $|\tilde{x}_m\ran$ $=$ $|x_0+m \eta\ran$. Moreover, in Eqs.\eqref{apoly}-\eqref{npoly},\eqref{qo1b},\eqref{qo2b} one has to replace $\eps \rightarrow \eta$. Such operators will be denoted as $\tilde{a}$, $\tilde{a}^\dag$, $\tilde{N}$ and $\tilde{b}$, $\tilde{b}^\dag$, $\tilde{N}$. Then, defining $|m\ran \ \equiv \ \frac{|\tilde{x}_m\ran}{([m]!)^\ha}$, we get
\bea
\tilde{a} |0\ran & = & 0 \, , \qquad |m\ran \ = \ \frac{(\tilde{a}^\dag)^{m}}{([m]!)^\ha} \, |0\ran \, , \\[2mm]
\tilde{a}^\dag \, |m\ran & = & [m+1]^\ha \, |m+1\ran \, , \,\, \tilde{a} \, |m\ran \ = \ [m]^\ha \, |m-1\ran \, , \\[2mm]
\tilde{N} \, |m\ran & = & m |m\ran \, ,
\eea
and similarly for the $b$ operators. The Hilbert space with basis $|m\ran$ is once more $\mathcal{H}_q$.

Then we have found the isomorphism\footnote{This is not a direct sum, because $\mathcal{H}^{{}_{n\leq 0}}_q \cap \mathcal{H}^{{}_{n\geq 0}}=\{\la |0\ran\}$.}
\be
\mathcal{H}_{Poly}^{x_0} \sim \mathcal{H}^{{}_{n\leq 0}}_q + \mathcal{H}^{{}_{n\geq 0}}_q \,,
\ee
valid for each fixed $x_0$. The operators $\{a, a^{\dag},N\}$ should be thought of as $\{a_{x_0}, a^{\dag}_{x_0},N_{x_0}\}$. Representations for different $x_0$ must be \emph{unitarily inequivalent}, in order to have the orthogonality of various polymer states for different $x_0$.
To understand this last delicate point, we look at Eq.\eqref{lqme2}. It is clear that, with the identification $\theta=k \eps$, and defining  $L=-Q/\eps=-i \pa_\theta$, $X_1= I_\eps = \cos \theta$ and $X_2=\eps \, P_\eps=\sin \theta$, we have a customary $e(2)$ Lie algebra:
\be
[L, X_1] = i  X_2 \, , \quad [L, X_2] = - i  X_1 \, , \quad [X_1, X_2] = 0 \,.
\ee
It is known that inequivalent representations of such algebra are related by the unitary \emph{improper} transformation \cite{Kastrup:2005xb}
\be \label{ude}
L_\delta \ \equiv \  G_{-\de}(\theta) \, L \, G_{\de}(\theta) \, , \quad G_\de(\theta) \ \equiv \ e^{i \de \theta} \, ,
\ee
and $\de \in [0,1)$, while $X_1, X_2$ take the same form. A basis of the Hilbert space $\mathcal{H}_\de$, is given by
\bea \label{emd}
e_{n, \de}(\theta) & \equiv & e^{i \theta (n+\de)} \, , \\[1mm] \label{scpr}
 (e_{n, \de}(\theta), e_{m, \de}(\theta)) & \equiv & \frac{1}{2 \pi} \int^{2\pi}_{0} e^{i (n-m)\theta} =\de_{n \, m} \, .
\eea
Identifying $\de \equiv x_0/\eps$ these coincide with $|x_n\ran$ (see Eqs.\eqref{psi2n},\eqref{psik0}).

The transformation \eqref{ude} corresponds to the one passing from $Q$ to $Q_{x_0}$, introduced in Eq.\eqref{qx0} to build the appropriate representation of $q$-WH. Then, $\mathcal{H}_{Poly}$ is decomposed into a direct sum of $\mathcal{H}_\de$. For each $\de$, such spaces are decomposed as $\mathcal{H}^{{}_{n\leq 0}}_q + \mathcal{H}^{{}_{n\geq 0}}_q$: \emph{then $\mathcal{H}_{Poly}$ is decomposed into an infinite sum of $q$-oscillator spaces}. Moreover, from the theory of $E(2)$ representations we know that rational values of $\de=n_1/n_2$, with $n_1,n_2 \in \mathbb{Z}$ correspond to representation of the $n_2$-fold covering of $E(2)$ Lie group, while irrational values correspond to representations of its universal covering \cite{Kastrup:2005xb}.

These last considerations represent a basis for the mathematical description of a quantum particle moving on a circle. Although a relationship between such system and PQM has been already pointed out \cite{PhysRevD.76.044016, Morales_T_cotl_2017,Berra_Montiel_2021}, here we want to emphasize the importance of $q$-calculus. One can write the complex map $z \equiv e^{-i \theta}=e^{-i k \eps}$, which realizes a compactification. Then, wavefunctions \eqref{psik0} are mapped into
\be
\psi_{\de}(z) \ = \ \sum^{+\infty}_{n=-\infty} \, c_{n} \, z^{n+\de}  \, .
\ee
A representation of the $q$-oscillator on such functions is given by\footnote{Note that, being $|z|=1$, $\dr/\dr z$ is always well-defined, even for $\de \neq 0$.}
\bea \label{qfbrde}
a_\de \, f(z) & = & \bar{D}^\de_q \, f(z) \  \equiv \   \frac{f(q z) q^{-\de}-f(q^{-1}z)q^{\de}}{(q-q^{-1})z} \, , \\[2mm]
a^\dag_\de \, f(z) & = &  z \, f(z) \, ,\ \ \ N_\de \, f(z)  =  z \, \frac{\dr f(z)}{\dr z}-\de f(z) \, ,  \label{nopde}
\eea
or, for $b$-modes
\bea \label{qfbr1de}
b_\de \, f(z) & = & D^\de_q \, f(z) \  \equiv \  \frac{f(q z) q^{-\de}-f(z)}{(q-1)z} \, , \\[2mm]
b^\dag_\de \, f(z) & = &  z \, f(z) \, , \ \ \  N_\de \, f(z) = z \, \frac{\dr f(z)}{\dr z}-\de f(z) \, .  \label{nop1de}
\eea
Note that $\bar{D}^\de_q z^{n+\de} \ = \ [n] z^{n+\de}$, $D^\de_q z^{n+\de} \ = \ [\![n]\!] z^{n+\de}$ and that $\bar{D}^\de_q,D^\de_q$ become the usual \emph{$q$-derivatives} for $\de=0$:
\bea
\label{qfbr}
\bar{D}_q \, f(z) &  \equiv &  \frac{f(q z)-f(q^{-1}z)}{(q-q^{-1})z} \, ,  \ \\[2mm] \label{qfbr1}
 D_q \, f(z) &  \equiv &  \frac{f(q z)-f(z)}{(q-1)z} \, .
\eea
In PQM it is customary to work in a fixed representation (i.e. with a fixed $x_0$) \cite{Morales_T_cotl_2017,Amelino_Camelia_2017,Berra_Montiel_2021}. A natural choice, which could be related to the time-reversal symmetry \cite{Kowalski:1998hx}, is $x_0=0$, i.e., precisely $\de=0$. In this case one recovers the usual Fock--Bargmann representation of the $q$-WH \cite{DamKul,klimyk2012quantum,Celeghini_1995,qSvN} and the natural framework is the $q$-calculus becomes. In fact, the scalar product is generally given by a \emph{Jackson integral} \cite{DamKul}
\be
(f,g) \equiv \int f^*(z) g(z) \dr_q \mu(z) \,,
\ee
but, as for us $|z| = 1$, this should be replaced by the standard scalar product of Eq.\eqref{scpr}, which does not depend from $\de$.

%\cite{klimyk2012quantum}
%%
%\bea \label{qfbr}
%a \, f(z) & = &  \bar{D}_q \, f(z) \  \equiv \  \frac{f(q z)-f(q^{-1}z)}{(q-q^{-1})z} \, , \\[2mm]
%a^\dag \, f(z) & = &  z \, f(z) \, , \ \ \ N \, f(z) \ = \ z \, \frac{\dr f(z)}{\dr z} \, ,  \label{nop}
%\eea
%%
%or \cite{klimyk2012quantum,Celeghini_1995,qSvN}
%%
%\bea \label{qfbr1}
%b \, f(z) & = &  D_q \, f(z) \  \equiv \  \frac{f(q z)-f(z)}{(q-1)z} \, , \\[2mm]
%b^\dag \, f(z) & = &  z \, f(z) \, , \ \ \ N \, f(z) \ = \ z \, \frac{\dr f(z)}{\dr z} \,.  \label{nop1}
%\eea
%
The importance of the $q$-calculus in PQM can be also appreciated from Eq.\eqref{lqme2}. Indeed, while in the usual momentum representation, $P$ acts as the multiplication by a standard number, $P f(k)=k f(k)$, in the present case
\be
P_q  f(k) \ \equiv \ P_{\eps(q)} f(k) \ = \  \frac{q-q^{-1}}{2 \log q} \, \lf[k\ri]  f(k) \,,
\ee
i.e. $P_q$ multiplies $f(k)$ by a $q$-number.
%%%%%%%%%%%%%%%%%%%%%%%%%%%%%%%%%%%%%%%%%%%%%%%%%%%%%%%%%%%%%%%%%%%%%%%%%%%%%
\section{Conclusions and Outlook}

We have shown here that the algebra $q$-WH is naturally represented in ${\cal H}_{Poly}$, and that the Hilbert space of PQM is then the sum of Hilbert spaces of the $q$-oscillator. In this way we offer here a \emph{natural}, although non-unique, solution to the problem of defining ladder operators in PQM through the ladder operators of the $q$-oscillator.

This might have far-reaching consequences, because it would strengthen the candidature of quantum groups as fundamental quantum structures of nature. In fact, reviving in a quantum gravity context the original spirit of the Leningrad school, our results suggest that quantum groups might lend their powerful and well-developed calculus as the \emph{natural calculus} of PQM in particular, and of quantum gravity in general. For instance, besides the expected impact on LQG, our results might be useful for quantum gravity theories based on discrete fundamental structures \cite{kleinert1987,Jizba:2009qf,Petruzziello:2020wkd}
%***CITARE Renate Loll, Discrete Approaches to Quantum Gravity in Four Dimensions, Living Reviews in Relativity***,
such as those based on finite-dimensional Hilbert spaces, see, e.g., \cite{ACQUAVIVA2017317, Acquaviva:2020prd}.

In particular, in the latter works \emph{topological inequivalence} among different mathematical representations/physical phases is important, and we have shown here that the inequivalence between superselected sectors of ${\cal H}_{Poly}$ is indeed of topological nature, as it stems from the relation of these representations with the representations of the Euclidean algebra $e(2)$ appearing in the quantization of a particle winding around in a circle. The latter correspondence also points to scenarios of condensed matter analog realizations via lattice crystals \cite{Kastrup:2005xb}.

Finally, Hopf algebras present many more operations than those introduced here, so we have not exploited their potential physical role. Actually, even the coproduct, that we have duly taken into account here, is not discussed from a physical perspective, as done instead in the applications to quantum fields in curved space of \cite{IORIO2001234}. In fact, much more could be extracted from these structures, especially in relation to quantum fields at finite temperature \cite{CELEGHINI1998455} and in turn to the description of spacetime horizons \cite{IORIO2001234}.

%%%%%%%%%%%%%%%%%%%%%%%%%%%%%%%%%%%%%%%%%%%%%%%%%%%%%%%%%%%%%%%%%%%%%%%%%%%%%%%%%%%
\section*{Acknowledgements}
A.I. and L.S. acknowledge support from Charles University Research Center (UNCE/SCI/013).
%%%%%%%%%%%%%%%%%%%%%%%%%%%%%%%%%%%%%%%%%%%%%%%%%%%%%%%%%%%%%%%%%%%%%%%%
%\section*{References}

%\bibliographystyle{apsrev4-1}
\bibliographystyle{apsrev4-2}
\bibliography{librarySvN}

\end{document}